\documentclass[]{spie}  
\usepackage[]{graphicx}

\title{Review of small-angle coronagraphic techniques in the wake of ground-based second-generation adaptive optics systems} 

\author{Dimitri Mawet\supit{a,b}, Laurent Pueyo\supit{c}, Peter Lawson\supit{b}, Laurent Mugnier\supit{d}, Wesley Traub\supit{b}, Anthony Boccaletti\supit{e}, John Trauger\supit{b}, Szymon Gladysz\supit{f}, Eugene Serabyn\supit{b}, Julien Milli\supit{a,g}, Ruslan Belikov\supit{h}, Markus Kasper\supit{i}, Pierre Baudoz\supit{e}, Bruce Macintosh\supit{j}, Christian Marois\supit{k}, Ben Oppenheimer\supit{l}, Harrisson Barrett\supit{m}, Jean-Luc Beuzit\supit{g}, Nicolas Devaney\supit{n}, Julien Girard\supit{a}, Olivier Guyon\supit{o}, John Krist\supit{b}, Bertrand Mennesson\supit{b},  David Mouillet\supit{g}, Naoshi Murakami\supit{p}, Lisa Poyneer\supit{j},  Dmitri Savransky\supit{j}, Christophe V\'erinaud\supit{g}, and James K.~Wallace\supit{b}
\skiplinehalf
\supit{a}European Southern Observatory, Alonso de C\'ordova 3107, Vitacura, Casilla 19001, Chile; \\
\supit{b}Jet Propulsion Laboratory, California Institute of Technology, Pasadena, CA 91109, USA;\\
\supit{c}JHU Department of Physics and Astronomy, 3400 N. Charles St, Baltimore, MD 21218, USA;\\
\supit{d}ONERA, Division Optique Theorique et Appliqu\'ee, BP 72, 92322 Chatillon Cedex, France;\\
\supit{e}LESIA, Observatoire de Paris, CNRS, 5 place Jules Janssen, 92195 Meudon, France;\\
\supit{f}Fraunhofer Institute, Gutleuthausstrasse 1, 76275 Ettlingen, Germany;\\
\supit{g}IPAG, 414 rue de la Piscine, Domaine Univ., BP 53, 38041 Grenoble Cedex 09, France;\\
\supit{h}NASA Ames Research Center, Moffett Field, CA 94035, USA;\\
\supit{i}European Southern Observatory, Karl-Schwarzschild-Stra$\ss$e 2, 85748 Garching, Germany; \\
\supit{j}Lawrence Livermore National Laboratory, 7000 East Ave, Livermore, CA 94550, USA;\\
\supit{k}NRC, Herzberg Institute of Astrophysics, Victoria, BC V9E 2E7, Canada;\\
\supit{l}American Museum of Natural History, New York, NY 10024, USA;\\
\supit{m}College of Optical Sciences, University of Arizona, Tucson, AZ 85721, USA;\\
\supit{n}Applied Optics Group, School of Physics, National University of Ireland, Galway, Ireland;\\
\supit{o}Subaru Telescope, National Astronomical Observatory of Japan, Hilo, HI 96720, USA;\\
\supit{p}Hokkaido University, Sapporo, Hokkaido 060-8628, Japan}

\authorinfo{Further author information: send correspondence to dmawet@eso.org}

\begin{document} 
  \maketitle 

\begin{abstract}
Small-angle coronagraphy is technically and scientifically appealing because it enables the use of smaller telescopes, allows covering wider wavelength ranges, and potentially increases the yield and completeness of circumstellar environment -- exoplanets and disks -- detection and characterization campaigns. However, opening up this new parameter space is challenging. Here we will review the four posts of high contrast imaging and their intricate interactions at very small angles (within the first 4 resolution elements from the star). The four posts are: choice of coronagraph, optimized wavefront control, observing strategy, and post-processing methods.  After detailing each of the four foundations, we will present the lessons learned from the 10+ years of operations of zeroth and first-generation adaptive optics systems. We will then tentatively show how informative the current integration of second-generation adaptive optics system is, and which lessons can already be drawn from this fresh experience. Then, we will review the current state of the art, by presenting world record contrasts obtained in the framework of technological demonstrations for space-based exoplanet imaging and characterization mission concepts. Finally, we will conclude by emphasizing the importance of the cross-breeding between techniques developed for both ground-based and space-based projects, which is relevant for future high contrast imaging instruments and facilities in space or on the ground.
\end{abstract}



\section{INTRODUCTION}
\label{sec:intro}  

\begin{figure}[!t]
\centerline{\includegraphics[width=16cm]{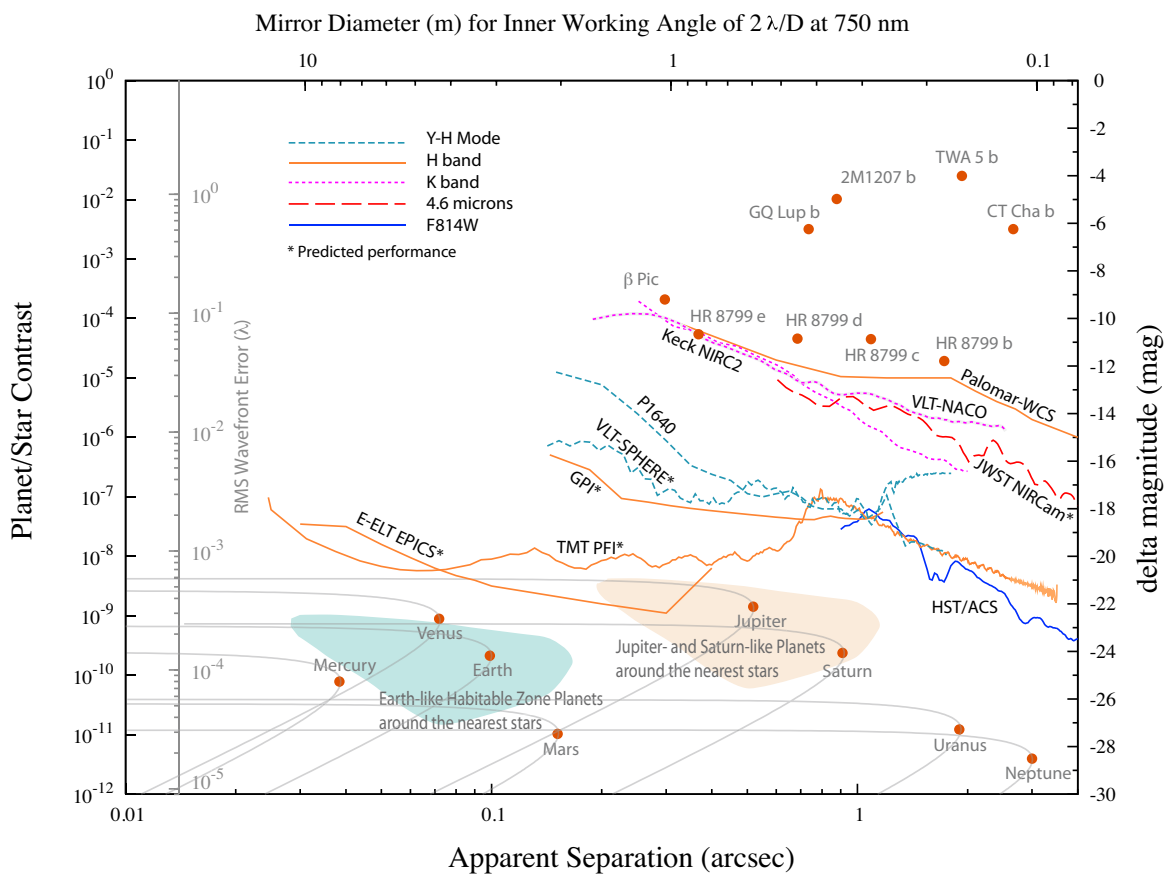}}
\caption{Snapshot of current and projected high contrast imaging capabilities in space and from the ground (see Ref.~Lawson et al.~2012, these proceedings, for additional details). The left axis shows both the Planet/Star contrast ratio together with the corresponding rms wavefront quality necessary to reach it (assuming a high order deformable mirror, e.g.~64 by 64 actuators). The right axis shows the corresponding $\Delta$ magnitude relative to the central star. The x axis shows the angular separation in arcsec. All detectivity curves are $5\sigma$ and scaled for a 1-hour observing time. An asterisk denotes predicted contrast for future instruments. Keck-NIRC2, VLT-NACO, Palomar-WCS, and HST-ACS curves show current representative capabilities of these high contrast imaging workhorse instruments. Improving upon these first generation instrument, we present here as well the expected progress of the second generation (e.g.~GPI and SPHERE) both in terms of contrast and inner working angle, which is the very focus of the current review. Note that the Palomar-P3K-P1640 second generation instrument is already on sky, and the curve presented here is based on real data. Longer term projections include JWST-NIRCam, TMT PFI, and E-ELT EPICS. Over-plotted are the K-band fluxes for 9 of the 15 or so extra-solar planets that have been imaged so far. In the lower part of the figure are plotted our solar system planets as they would appear in reflected light around a Sun-like star at a distance of 10 pc. One caveat to this comparison plot is the diversity in wavelength ranges covered, and the obvious but inevitable overlap between the reflected light and thermal emission regimes.
\label{fig0}}
\end{figure}

High contrast imaging of extra-solar planetary systems provides a complete toolkit to characterize planets and their host system \cite{Absilmawet2010} through, e.g., orbital motion \cite{Soummer2011, Chauvin2012}, spectro-photometry of planetary atmospheres \cite{Janson2010, Galicher2011, Bonnefoy2011}, or planet-disk interactions \cite{Lagrange2012}. However, imaging extra-solar planets around other stars constitutes a multiple challenge, and the practical hurdles are numerous. First of all, the angular separation between planets and stars is very small (e.g.~$<$100 mas for a 1-AU distance at 10 pc), usually requiring diffraction limited capabilities on ground-based 8-meter class telescopes in the near-infrared\footnote{Note the exception in Ref.~\citenum{2010Natur.464.1018S}, which presented a snapshot of 3 out of the 4 planets of HR~8799\cite{2010Natur.468.1080M} taken with an adaptively-corrected 1.5-meter segment of the 5.1-meter Hale telescope of Palomar observatory, and a next-generation vector vortex phase-mask coronagraph\cite{2005ApJ...633.1191M}.} or space-based 2-meter class telescopes in the visible. Second, the contrast between a planet and its host star ranges from $\simeq 10^{-3}$ for hot giant planets in the infrared to $\simeq 10^{-10}$ for Earth-like planets in the visible (Fig.~\ref{fig0}). The contrast issue demands exquisite image (hence wavefront) quality to feed coronagraph devices, most of the time very specialized observing strategies (e.g., angular differential imaging or ADI\cite{Marois2006}), and corresponding data reduction techniques such as the Locally Optimized Combination of Images (LOCI\cite{Lafreniere2007}). 

Fig.~\ref{fig0} showcases an up-to-date overview of current and projected high contrast imaging capabilities, highlighting workhorse instruments such as NAOS-CONICA \cite{2010SPIE.7736E..89G} (NACO hereafter), the adaptive optics near-infrared camera of the Very Large Telescope (VLT), NIRC2, its alter-ego of the Keck telescope\cite{2003SPIE.4841....1M}, the Palomar Well-Corrected Subaperture (WCS\cite{2007ApJ...658.1386S}), and the HST Advanced Camera for Surveys (ACS). The plot also shows projected contrast capabilities of the major next generation ground-based high contrast imagers, namely the Gemini Planet Imager\cite{2008SPIE.7015E..31M} (GPI) and VLT-SPHERE \cite{2006Msngr.125...29B}, together with the Thirty Meter Telescope Planet Finder Instrument (TMT PFI\cite{2006SPIE.6272E..20M}), the European-Extremely Large Telescope Exoplanet Imaging Camera and Spectrograph (E-ELT EPICS\cite{2010SPIE.7735E..81K}), and the James Webb Space Telescope (JWST) near IR Camera (NIRCAM\cite{2004SPIE.5487..628H}). Note that the JWST mid IR imager (MIRI\cite{2004SPIE.5487..653W}) will provide similar working angles and detection limits in terms of mass at longer wavelengths, which partially compensates for its lower contrast\cite{2005AdSpR..36.1099B}. Finally, let us emphasize the recent ground-breaking demonstration of the Palomar P3K-P1640\cite{2011PASP..123...74H} second-generation instrument (extreme AO and integral field spectrograph), which is already on sky and nearing in practice the contrast levels only projected for its competitors (Oppenheimer et al.~2012, these proceedings).

Digging deep and close in requires optimized techniques, that can be extensions of the methods used at larger angles, but are most of the time entirely dedicated, because the latter are not adapted or even fail. The difficulty of small-angle high contrast imaging is daunting, but the prize interesting both technically and scientifically, as we will show in this review. The preferred metric for small-angle coronagraphy is called the inner working angle (IWA). The IWA is universally defined as the 50\% off-axis throughput point of a coronagraphic system, most of the time expressed in terms of the resolution element ($\lambda/d$). As we will detail below, providing access to small IWA enables the use of smaller telescope, allows covering wider wavelength range, and increasing the yield and completeness of circumstellar environment (planets and disks) surveys and characterization campaigns. 

Here we propose to review what we consider as the four pillars of high contrast imaging at small angles, and then derive the lessons learned from existing facilities and upcoming ones. Before going further, the authors would like to stress the limited scope of this review, which was intentional. Indeed, we will focus in the following on very small angles, from $\simeq 1 \lambda/d$, which is considered as close to the fundamental limits of coronagraphy\cite{Guyon2005a}, and up to  $\simeq 4 \lambda/d$, where more classical techniques as the ones exposed below can be very efficiently used. Note that within the resolution element $\lambda/d$, interferometric techniques such as sparse aperture masking\cite{2011A&A...532A..72L} or single pupil nulling interferometry \cite{2011ApJ...736...14M} provide the necessary coverage complementary to classical coronagraphy. 

\begin{figure}[!t]
\centerline{\includegraphics[width=16cm]{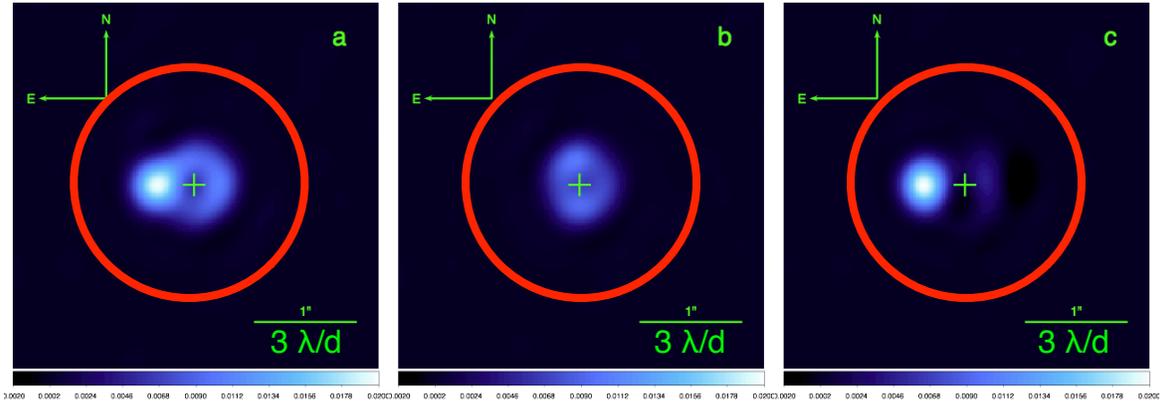}}
\caption{Image of $\epsilon$ Cephei and its close companion discovered recently using a vector vortex coronagraph behind the 1.5 m WCS at Palomar\cite{2010ApJ...709...53M,2011ApJ...738L..12M}. $a=$target image, $b=$reference star, $c=b-x\times a$. The candidate companion is 50 times fainter than $\epsilon$ Cephei, and lies at an angular separation of 330 mas, or 1.1 $\lambda/d$ for the WCS, making it the smallest angle detection ever realized with a coronagraph in terms of $\lambda/d$ units\cite{2011ApJ...738L..12M}. The red circle represents the $3\lambda/d$ inner working angle of typical new generation Lyot coronagraphs (e.g.~the apodized Lyot coronagraph). While the Lyot coronagraph family still presents many advantages, it appears clearly in this example that even reduced contrast capabilities within the blind spot of Lyot coronagraphs is scientifically undoubtedly important.
\label{fig1}}
\end{figure}

\section{Rationale of high contrast imaging at small angles}
\label{sec:science}  

Accessing very small angles has several advantages. For a given telescope size, enabling small IWA observations opens up a new parameter space:
\begin{enumerate}
\item On one hand, for a given distance, a smaller IWA allows getting closer to the central star, unveiling a new search area proportional to $IWA^{-2}$ (ignoring inclination effects). Probing the inner regions of stellar systems is crucial to understand formation and evolution of rocky planets within the primordial and debris disk, respectively. This is also where the habitable zone is located for the vast majority of nearby systems (1 AU at 10 pc is $0."1$ or 2$\lambda/d$ for a 2-meter telescope in the V band). Moreover, enabling smaller IWA in the visible regime allows benefiting from increased reflected light luminosity (under the hypothesis that the albedo is constant, which is supposedly not the case within 1 AU for eccentric orbits\cite{2010ApJ...724..189C,2012A&A...541A..83M}), which eases the constraints for both the detection and subsequent characterization.
\item On the other hand, neglecting sensitivity at first order, for a given angular separation $r$, a smaller IWA increases the volume of accessible objects by a significant $IWA^{-3}$, enabling the search for planets in more distant star forming regions (Mawet et al.~2012, in press), which constitutes very attractive but challenging reservoirs of targets to survey. Indeed, most of the objects imaged so far are orbiting young stars (see exoplanet.eu for a thorough and up-to-date list). Youth is the current bias of high contrast imaging, as short period, inclination or distances (orbital and/or parallactic) are the biases of radial velocity, transit and micro-lensing techniques, respectively. The thermal radiation of young exoplanets peaks in the near-infrared, making them more easily detectable by several orders of magnitude than if we were to observe them in reflected light in the visible. Since the detected emission comes from the intrinsic thermal radiation of the planet, its physical properties (temperature, mass and radius) can only be inferred based on cooling track models, which critically depend on age and formation mechanisms/history\cite{2007ApJ...655..541M}. Probing reservoirs of systems with a common history is then crucial to advance our understanding of these evolutionary models. 
\end{enumerate}

In terms of the resolution element $\lambda/d$, where $\lambda$ is the observing wavelength, and $d$, the diameter of the telescope's primary mirror, getting down to this physical scale enables the use of smaller telescopes\cite{Serabyn2011a}, which provides a substantial discount on a dedicated space-based project. Of course, the discount comes at a price in terms of substantially longer integration times for a given SNR (hence tighter requirements on stability) and substantially higher sensitivities to low-order wavefront errors and alignment drift (see Sect.~\ref{subsec:wfs}).
On the wavelength range aspects, gaining access to small angles enables pushing the limits to longer wavelengths (e.g.~L, M, N bands, i.e.~from 3 to 20 $\mu$m) without suffering from the decrease in effective IWA. This last point is of crucial importance for the extremely large telescopes, where thermal infrared high contrast imagers are foreseen to open up the unique parameter space of high sensitivity/high angular resolution long wavelength imaging from the ground\cite{2008SPIE.7014E..55B}. 

In summary, small IWA capabilities provide significant scientific gains and technical interests. One important note about accessing small IWA is that, even if contrast performances (hence sensitivity in the detectability sense of the word) are reduced, having the capability of imaging in these regions as opposed to being totally blind, is a must-have! In the example shown in Fig.~\ref{fig1}, the close companion to $\epsilon$ Cephei was imaged at 1.1 $\lambda/d$. Even though the contrast involved in this example is mild, this detection clearly demonstrates the scientific potential of small IWA coronagraphy\cite{2011ApJ...738L..12M}.

\begin{figure}[!t]
\centerline{\includegraphics[width=15cm]{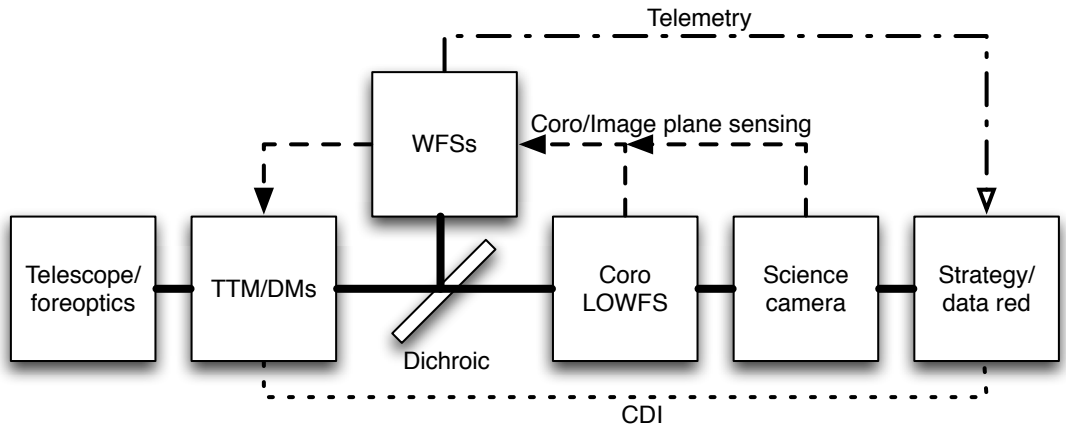}}
\caption{Block diagram of a typical small-angle coronagraphic instrument, showing the control systems (tip-tilt mirror or TTM, and deformable mirrors or DMs), the main wavefront sensor (WFS) critical on ground-based telescopes, the coronagraph, and associated low-order wavefront sensor (LOWFS). The science camera itself is an essential part of the system, since it allows calibrating the residual non-common path aberrations (NCPA) unseen by other sensors. Note that observing strategies and data reduction techniques now play more and more significant roles. They can make also make good use of telemetric data coming from various sensors, passively. Another promising approach to detection is the use of dynamic coherent modulation of the signal (e.g., with the DM), a technique known as coherent differential imaging (CDI, see Sect.~\ref{subsec:strategy}).
\label{fig2}}
\end{figure}

\section{The four pillars of high contrast imaging at small angles}
\label{sec:fourpillars}  

In this section, we present the four pillars of high contrast imaging at very small angles, i.e.~between $\simeq 1$ and $\simeq 4 \lambda/d$. As mentioned earlier, we decided to focus on this angular separation range to limit the scope of this review and restrict ourselves to the very specific techniques that are needed in this small-angle regime. The mid-to-large separation (beyond 4 $\lambda/d$) has extensively been covered in the literature for the past ten years. As we will show, small IWA imaging demands dedicated techniques all working in concert. Fig.~\ref{fig2} presents a block diagram of a typical small-angle system. The key point is the multiplicity of the sensing devices, all meant to stabilize the PSF on the coronagraph (and science camera). As we will show later, pointing stability and control over low-order aberrations is indeed essential, because most forms of their subsequent leakage through the coronagraph will mimic the signal of either potential companions or off-axis extended sources.

\subsection{small-angle coronagraphs}
\label{sec:fourpillars}  

\begin{figure}[!t]
\centerline{\includegraphics[width=16cm]{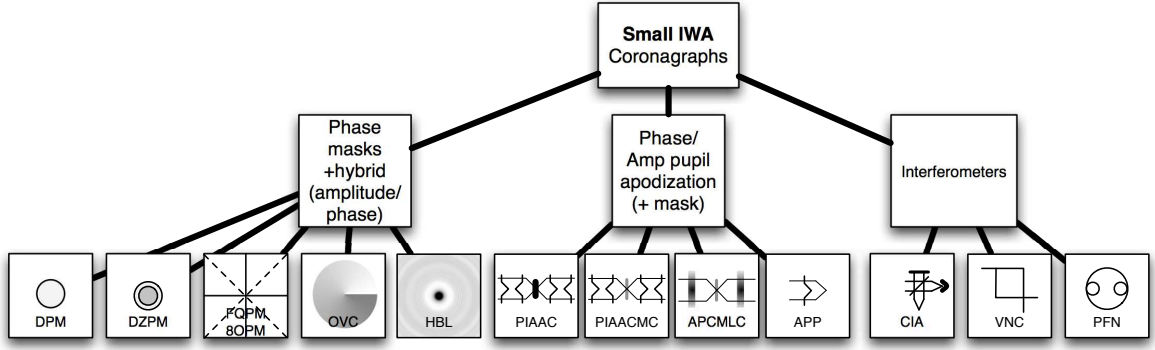}}
\caption{The family tree of small-angle coronagraphs (representative cases included). DPM: disk phase mask. DZPM: dual zone phase mask. FQPM/8OPM: four-quadrant phase mask, 8 octant phase mask. OVC: optical vortex coronagraph. HBL: hybrid band-limited. PIAAC: phase induced amplitude apodization. APP: apodizing phase plate. CIA: ``coronagraphe interferentiel achromatique''. VNC: visible nuller coronagraph. PFN: Palomar fiber nuller. Common to all these concepts is some sort of phase manipulation, either in the focal plane or in the pupil plane, or both. Three main families can be pointed out: phase masks, phase/amplitude pupil apodization + focal plane masks (phase and/or amplitude), and interferometers.
\label{fig3}}
\end{figure}

Only a dozen of coronagraphs are transmitting within $1-4\lambda/d$, which can already be considered as a healthy portfolio. Fig.~\ref{fig3} and Table~\ref{tab:smalliwacoro} present a tentative census of these particular devices. Not all of them possess the same level of technological maturity, but they all without exception have been the subject of significant developments and laboratory demonstrations. Table~\ref{tab:smalliwacoro} summarizes the working principles of these coronagraphic systems in terms of pupil/focal plane phase/amplitude action.

Along with the  ``coronagraphe interferentiel achromatique'' (CIA\cite{2000A&AS..141..319B}), which is a mono-pupil interferometer, the disk phase mask (DPM), also known as the nulling coronagraph\cite{1999PASP..111.1321G}, was one of the first coronagraphs proposed to tackle the blindness of the classical Lyot coronagraph at small angles\footnote{The classical and more recent apodized version of the Lyot coronagraph are typically limited to $3-4\lambda/d$ inner working angle, mostly because they rely on amplitude manipulation to induce the attenuate starlight diffraction, covering a large area of the PSF at the coronagraph plane. Note that these new Lyot coronagraphs can in principle be optimized to access small IWA, but at the expense of overall throughput.}. The DPM was the first coronagraphic mask to introduce image plane phase modulation in an attempt to null the light in a downstream pupil plane. Indeed, by phase shifting the core of the PSF by $\pi$ radian over a radius of about $0.53\lambda/d$, the superposition of the diffracted waves corresponding to both parts of the divided PSF, would almost perfectly cancel out in the downstream pupil. Phase modulation of the PSF thus enabled the use of transparent masks that are conveniently letting more  of the off-axis signal through them. The IWA of the DPM would be close to the diffraction limit. However, the DPM was plagued by two major weaknesses related to chromatic effects: chromaticity of the $\pi$ phase shift and of the fixed radius of the phase mask vs the wavelength-dependent radius of the PSF. Chromaticity is undoubtedly the Achilles heel of this new family of coronagraphs, even though multiple solutions have been worked out over the years. For instance, the four-quadrant phase-mask (FQPM\cite{2000PASP..112.1479R}) was especially invented to permanently solve the radial chromaticity of the DPM by introducing azimuthal phase modulation to replace the radial modulation of the DPM. Later on the optical vortex coronagraph (OVC\cite{2005ApJ...633.1191M}) extended this concept further by making the azimuthal modulation completely smooth. In the meantime, the dual-zone phase mask was introduced (DZPM\cite{2003A&A...403..369S}) to compensate for the phase and size chromaticism of the DPM by adding complexity (degrees of freedom) to the radial modulation.

The concept of loss-less pupil apodization through phase remapping, or phase-induced amplitude apodization (PIAA\cite{2003A&A...404..379G,2005ApJ...622..744G}) was introduced at about the same time as the wave of new phase masks. It is remarkable as it provides ultimate theoretical answers to both the IWA and throughput drawbacks of classical apodized coronagraphs \cite{2005ApJ...618L.161S}. The family of phase pupil remappers is also quite large and not limited to the specific PIAA development. For instance, let us emphasize the successful implementation of the apodizing phase plates (APP) at the MMT\cite{2006SPIE.6269E..55C} and NACO\cite{2007ApJ...660..762K}.

Finally, note that the hybrid band-limited (HBL) coronagraph makes use of both amplitude and phase modulation\cite{2011SPIE.8151E..14T}, which allows it to also access smaller angles ($1\lambda/d$ is in theory feasible with a second-order HBL, but at the expense of significant throughput losses). The HBL coronagraph currently holds the world record of the deepest and broadest (in terms of bandwidth) demonstrated contrast level in the lab, down to an IWA of $3 \lambda/d$ (Sect.~\ref{sec:stateoftheart}). 

\begin{table}[!t]
\caption{Small IWA coronagraphs fundamental recipes in terms of phase ($\phi$) / amplitude ($A$), pupil (upstream and downstream from the focal plane mask) / focal plane modulation formula. See Fig.~\ref{fig3} for the acronym definitions. LS stands for Lyot stop.} 
\label{tab:smalliwacoro}
\begin{center}       
\begin{tabular}{|l|l|l|l|l|} 
\hline
Name &Th. IWA ($\lambda/d$) &Upstream Pupil &Focal plane &Downstream pupil\\
\hline
DPM &1 &N/A &$\phi=f(r)$  &LS\\
DZPM &1.5 &$A=f(r)$ &$\phi=f(r)$  &LS\\
FQPM &1 &N/A &$\phi=f(\theta)$  &LS\\
8OPM &1.7 &N/A &$\phi=f(\theta)$  &LS\\
OVC2 &1 &N/A &$\phi=f(\theta)$  &LS\\
OVC4 &1.7 &N/A &$\phi=f(\theta)$  &LS\\
HBL2 &1 &N/A &$A,\phi=f(r)$  &LS\\
HBL4 &2.5 &N/A &$A,\phi=f(r)$  &LS\\
PIAA &1 &$\phi => A =f(r)$ &$A=f(r)$  &$\phi => A=f^{-1}(r)+$LS\\
PIAACMC &0.7 &$\phi => A=f(r)$ &$A,\phi=f(r,\theta)$  &$\phi => A=f^{-1}(r)+$LS\\
APP &2 &$\phi=f(r)$ &N/A  &N/A\\
\hline 
\end{tabular}
\end{center}
\end{table} 

\subsection{Wavefront control of low-order aberrations}
\label{subsec:wfs}  

Because of their intrinsic small IWA, all the coronagraphs above are inevitably sensitive to low-order aberrations (tip-tilt, focus, astigmatism, coma, etc.), which leak directly around the mask center to mimic off-axis signal. Symmetric leakage, for very small asymmetric aberrations or symmetric aberrations (e.g.~focus), mimics signal from a putative circumstellar disk while asymmetric leakage, such as with larger decentering, mimics off-axis companions (residuals in Fig.~\ref{fig1} are the clear demonstration of the detrimental effects of low-order aberrations leakage).  

\begin{figure}[!t]
\includegraphics[width=8.5cm]{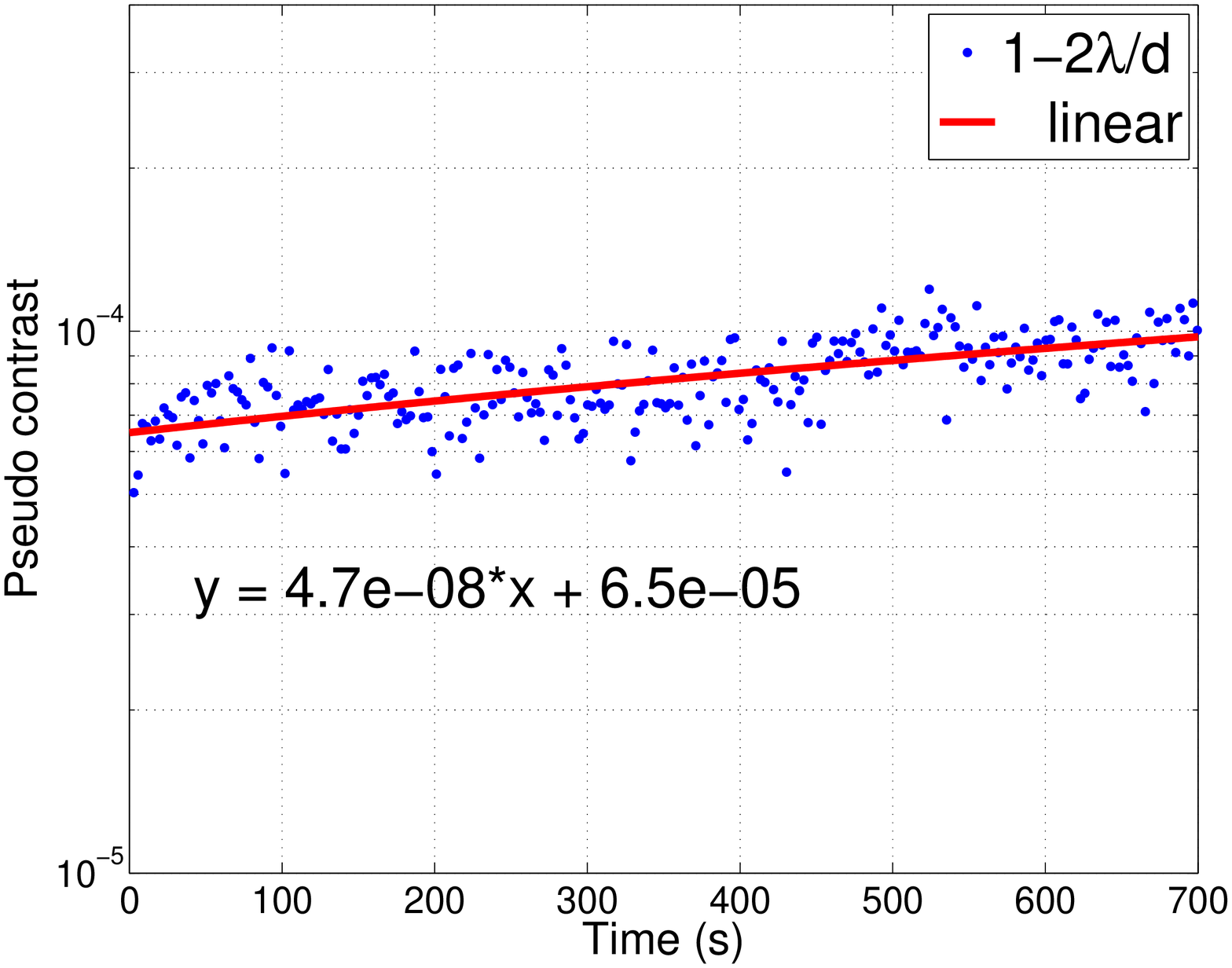}
\includegraphics[width=8.5cm]{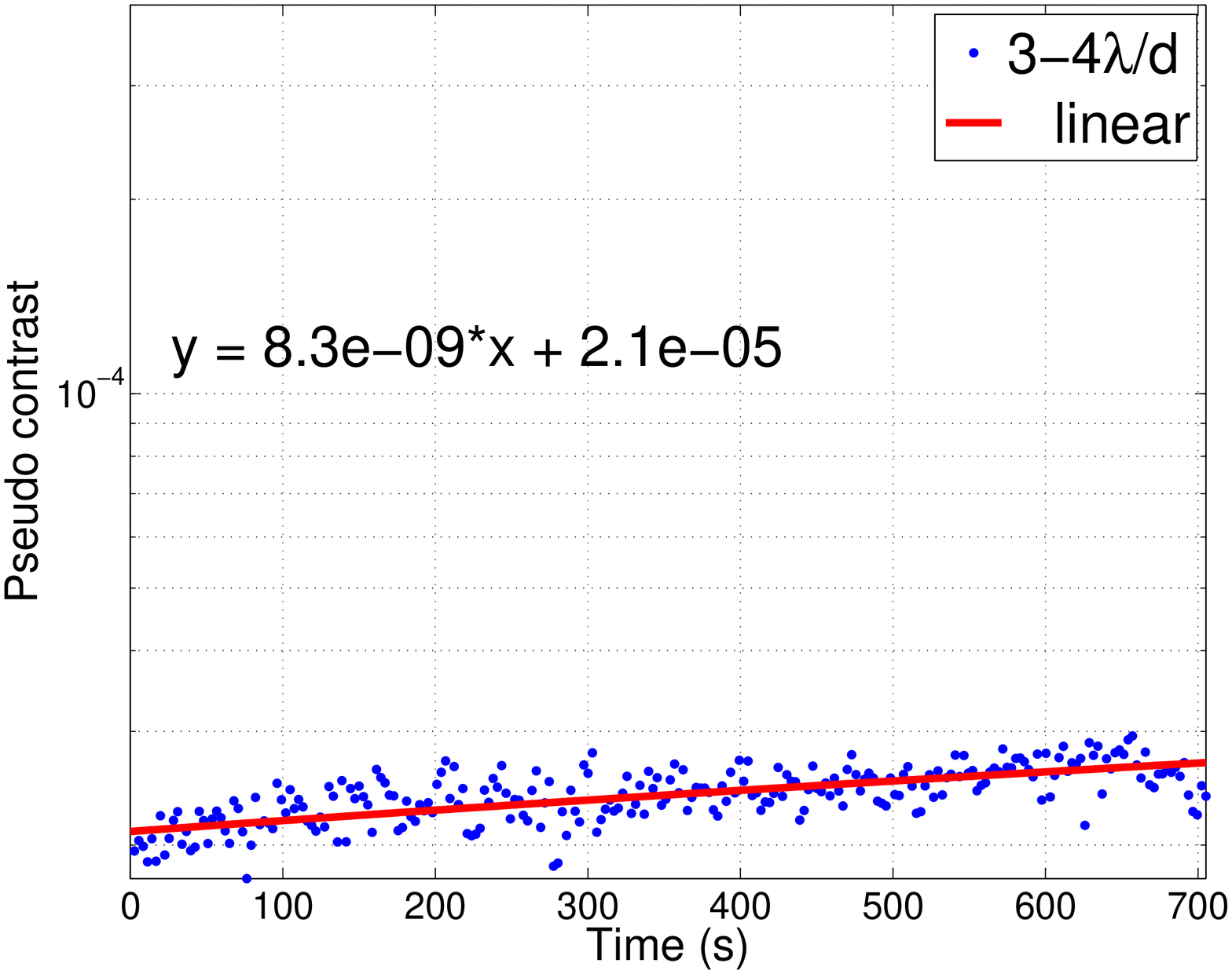}
\caption{Contrast variability for a ``zeroth-generation'' extreme AO system\cite{2010Natur.464.1018S,2010ApJ...709...53M} at very small angles (1 to 4 $\lambda/d$). As expected, contrast degrades with time, increasing toward smaller angles. In this particular example (1.5-meter well-corrected subaperture), the rate of contrast degradation due to speckles decorrelation approaches $5\times 10^{-8}$ per second at 1 $\lambda/d$, reflecting the performance of the pointing loop control which was here implemented manually (with feedback from the science image).
\label{fig4}}
\end{figure}

One obvious way to prevent the leakage in the first place is to control these aberrations down to equivalent contrast levels set by the science requirements. Of course, this is easier said than done, mostly because measuring these aberrations with the needed accuracy is very difficult and requires high dynamical range as well. Another reason for the difficulty of the task of low-order aberration control, as opposed to higher-order aberrations, is that the timescales involved are much shorter. At the extreme end of the spectrum, the tip-tilt error is probably the most difficult to control in real time because it is usually occurring very rapidly, due to e.g., seeing residuals and/or vibrations\cite{2010JOSAA..27A.122M}. In Fig.~\ref{fig4}, we show the contrast degradation due to low-order speckle decorrelation as a function of time, as measured for the zeroth-generation extreme adaptive optics system of the 1.5-meter Well-Corrected Subaperture (WCS) of the 5.1-meter Palomar's Hale telescope\cite{2006SPIE.6272E..91S}, using a vector vortex coronagraph. The relevant figure in these plots is the rate of contrast degradation due to speckle decorrelation, measured in pseudo-contrast units per second. As expected, the degradation rate is fast at very small angles ($1-4 \lambda/d$), i.e.~on the order of $1-5\times10^{-8} / s$, and accelerating when closing in towards the central star (which is partly explained by their increasing intensity towards the center as well). Indeed, we measured in this particular configuration about 5 times as fast a decorrelation rate at $1-2 \lambda/d$ than at $3-4\lambda/d$, pointing to the absolute necessity of controlling very low-order aberrations as swiftly as possible. Note that fast tip-tilt control was in this case only optimized by varying the gains of the tip-tilt loop to reduce apparent jitter, which has a characteristic doughnut shape through the vortex coronagraph (its size increases with the tip-tilt residuals so minimizing the doughnut diameter is a good visual way of reducing the jitter and optimizing the control loop gains). On the other end of the spectrum, slow pointing variations were adjusted in open loop with an automatic drift compensation. Residuals (on the order of $0.01\times\lambda/d$) were then compensated by hand on a minute basis, using the science image itself.

\begin{figure}[!t]
\centerline{\includegraphics[width=14cm]{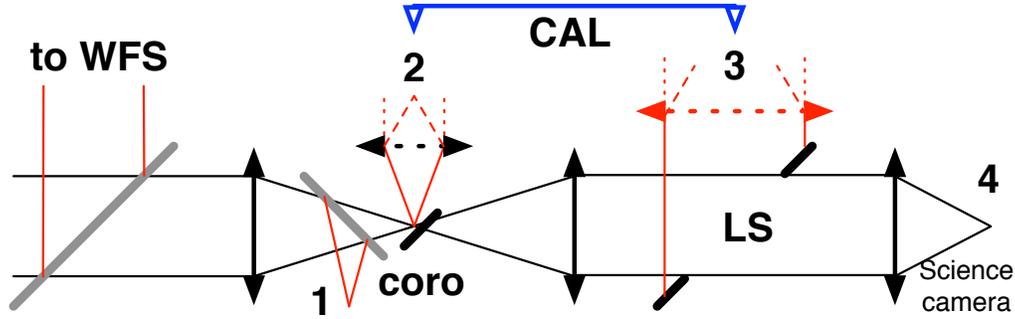}}
\caption{Illustration of the variety of low-order aberrations sensing schemes proposed, considered or implemented in current and next-generation systems. The philosophy is to provide a quasi real-time measurements of low-order aberrations (tip-tilt, focus, ...) that is affecting the coronagraphic plane (and not the final camera!). Despite the usual WFS which senses most of the atmospheric tip-tilt signal (with various efficiencies as discussed later), there are 4 other possible locations (see the text for additional details about each location).
\label{fig5}}
\end{figure}

Guided by the lessons of this particular example and other similar experiments, we identified three factors that need to be included in the trade-off while designing a low-order wavefront sensor, ordered here by their practical importance:
\begin{enumerate}
\item Location: several locations for the sensing of low-order aberrations are possible (see Fig.~\ref{fig5}), but as a general rule, the closer to the coronagraph and/or science camera, the better. Indeed, in coronagraphically-assisted high contrast imaging, the stability is primarily required at the coronagraph level to ensure a constant attenuation profile and thus behavior of the residual speckle field. In order to avoid non-common path interference, the measuring apparatus needs to be as simple as possible and as close as possible to the coronagraph.  Location 1, which has been adopted by the differential tip-tilt sensor (DTTS\cite{Baudoz2010b}) of SPHERE, is just upstream from the coronagraph, and makes use of a dichroic sending a fraction of the PSF signal to a dedicated fast detector. The advantage is the proximity to the coronagraph, and relative simplicity of the control algorithm. Drawbacks include the calibration to the coronagraph centering, and amplitude errors generated by the dichroic, which is located in the worst location for Talbot phase-amplitude conversion propagation effects. Location 2 uses the reflection of the coronagraphic spot, as in the so-called coronagraphic low-order wavefront sensor (CLOWFS) tailored to the PIAA coronagraph\cite{2009ApJ...693...75G}. Note that similar schemes have also been proposed, studied and even implemented for the reflection off or transmission through a Lyot coronagraph central region. The latter solution has been adopted for the apodized Lyot coronagraphs of P1640 and GPI\cite{2010lyot.confE..51V,2010SPIE.7736E.179W}, and is the preferred solution for the HBL as well. The advantage here is the extreme proximity to the occulting spot. The drawback of this solution is that it can only be implemented around coronagraphic masks that reflect enough starlight (more difficult for phase-mask coronagraphs). Location 3 redirects part of the light to a dedicated sensor, such as a Shack-Hartman sensor as for GPI LOWFS. The advantage of this solution is simplicity. The disadvantage of the SH sensor is its relative inefficiency at detecting small low-order aberrations\cite{2005MNRAS.357L..26V}. The final location is at the science camera itself, which is a privileged location to sense non-common path aberrations upstream, but also downstream from the coronagraph (Sauvage et al.~2012, these proceedings). However, the signal has been convolved by the coronagraph, so that its interpretation requires additional interpretation and modeling, but also provides enhanced dynamic range for the pointing error signal (e.g.~the residual doughnut of the vortex coronagraph, which size is indicative of the amplitude of the residual jitter). 
\item Sensitivity: since the low-order aberrations are the ones that vary the fastest, and since stars are faint, there are only so many photons once gets per mode. Ideally, one would want to get as many photons as possible from the low-order modes in order for the control loop to run fast. There is another degree of complication: to be able to run fast, one has to compute the ``sensing to control'' commands fast. If the errors are small then the inversion is straightforward as it is just a straight up linear inversion. If the errors are large, then one enters the non linear regime, with the consequence of having to go through a full phase retrieval calculation, which might still be tractable numerically given how fast computers go now, and which brings us to the third trade-off item.
\item Computational cost: driven by the complexity of the physics behind the chosen LOWFS scheme and the needed temporal bandwidths.
\end{enumerate}

Table~\ref{tab:wavefrontctrl} summarizes the main techniques recently developed to provide the necessary information to control the low-order aberrations with the tip-tilt and deformable mirrors. Typical performance levels range from 0.1 to 1 mas rms, which are sufficient to bring tip-tilt leakage under control in the visible and near-infrared regime, respectively.

\begin{table}[!t]
\caption{Wavefront control techniques specifically tailored to small angles.} 
\label{tab:wavefrontctrl}
\begin{center}       
\begin{tabular}{|l|l|l|l|l|l|}
\hline
Name &Example &Sensing method &Location &Perf.~($\lambda/d$) \\
\hline
LOWFS &GPI &SH+CAL\cite{2010SPIE.7736E.179W} &Behind coro (2-3) &$2\, 10^{-3}-5\, 10^{-2}$ (Ref.~\citenum{2010SPIE.7736E.179W} )\\
DTTS &SPHERE &Centroid &Upstream coro (1) &$\simeq 2\, 10^{-3}$ (Ref.~\citenum{Baudoz2010b})\\
Image plane &SPHERE &Table/model &Science camera (4) &$\simeq 6\, 10^{-2}$ (Ref.~\citenum{Mas2012})\\
CLOWFS &ScEXAO &Int.~matrix &Refl. light from occ. (2)  &$\simeq 1\, 10^{-3}$ (Ref.~\citenum{2009ApJ...693...75G})\\
\hline
\end{tabular}
\end{center}
\end{table} 

\subsection{Observing strategy}
\label{subsec:strategy}  

While angular differential imaging\cite{Marois2006} (ADI) and spectral differential imaging\cite{2000PASP..112...91M} (SDI) have proved to be or are expected to be very efficient at larger angular separation, these modulation techniques can not be applied efficiently at small angles. 

Indeed, for ADI, the linear displacement of the putative companion being proportional to $r\theta$, and given the typical $1-\lambda/d$ exclusion cone necessary to avoid self-subtraction, the amount of parallactic angle modulation necessary to meet this requirement becomes very constraining. Typically, at $2\lambda/d$, a displacement of $1\lambda/d$ requires an PA rotation of $> 25^\circ$, which is typically achieved in about 1 hour (the amplitude of PA rotation depends on the target coordinates and is maximum for an observation when the star passes the meridian). This time delay would prevent an adequate level of speckle correlation, and would thus render any subsequent speckle subtraction inefficient. Also for extended objects, artifacts due to self-subtraction are more likely to happen as the IWA gets smaller (Milli et al.~2012, in press).

The same reasoning goes for the multiple-channel SDI (with an integral field spectrograph\footnote{Here we make the distinction between multiple-channel SDI, which uses the radial modulation of speckles in multiple images (datacube), and dual-channel/dual-beam SDI, which only uses two images, assuming that the signal is present in one, and absent or significantly attenuated in the other. To avoid any confusion between the two different techniques, we use SDI to refer to the multiple-channel/radial modulation SDI, and DBI to refer to the dual-channel technique.}), but in the radial direction. A full $\lambda/d$-wide speckle elongation at an angular separation of $2\lambda/d$, would require a spectral coverage of 50\%, almost consuming the full range of existing and planned integral field spectrograph. Again, at such extreme of the modulation scale, decorrelation of the speckles is very likely (due to chromatic propagation effects, for instance), making multiple-channel SDI very inefficient for small angles as well.

Dual-beam imaging (DBI) makes two simultaneous images while using a physical property of the off-axis object that supposedly has a signature when differentiating the two images. For instance, DBI exploits intrinsic spectral features such as $CH_4$\cite{2006ApJ...641L.141B}, or polarization (PDI\cite{2009ApJ...701..804H}). Therefore, it is not affected by geometrical effects described above for, e.g.~multiple-channel SDI. Therefore, this technique would also work at very small angles but assumes that the off-axis signal presents significant molecular lines or degree of polarization, which still needs to be verified!

For these reasons, reference star differential imaging (RDI) is the preferred and well-proven solution. As was recently demonstrated\cite{2010Natur.464.1018S,2010ApJ...709...53M}, it is expected that the stability of next generation systems will be such that a high degree of correlation will persist between the target and the reference stars (see Fig.~\ref{fig4}). The stability can only be exploited to use RDI efficiently, if and only if three other important constraints can be met:
\begin{enumerate}
\item Duty cycle: if the duty cycle between their respective observations can be reduced to a few minutes, very high contrast can be expected from the resulting calibration (Fig.~\ref{fig4}).
\item Magnitude and color matching between the target and its references: the visible and near-infrared magnitudes need to be the same to ensure similar AO and various loops behavior, and comparable SNR of the field of quasi-static speckles, respectively.
\item Parallactic angle matching: to avoid flexure-induced semi-static speckle variations as much as possible and to ensure a consistent telescope orientation with respect to the instrument between the target and the reference, the calibration stars have to be chosen and observed at the same parallactic angle as the target star.
\end{enumerate}

This set of constraints is significantly limiting in terms of sky coverage, and involves careful planning and execution of the observations. These conditions can only be met on a best effort basis since the availability of a suitable reference fulfilling the set of constraints is never guaranteed, which is one of the drawbacks of the reference star differential imaging strategy. Finally, we note that the success of RDI at small angles will greatly depend on the performance of the LOWFS. If the LOWFS works well on both the target and the reference stars, and the residual noise is only due to the photon noise at the LOWFS detector, then RDI will perform ideally. However, if  there are systematics in the LOWFS, then the residuals will be star dependent. In this case telemetry and telescope history will even be very important to understand the systematics affecting the system. 

The statistical properties of speckles is a potential way of discriminating between speckles and real off-axis objects. This method would also work at very small angles. In Ref.~\citenum{2010JOSAA..27A..64G}, it is shown that with the modified Rician distribution one cannot describe the statistics of light on axis. A dual solution is proposed: the modified Rician distribution for off-axis speckle and gamma-based distribution for the core of the point spread function. From these two distributions an optimal statistical discriminators between real sources and quasi-static speckles can be derived.

Another promising approach, especially when all other methods fail, is the so-called ``coherent'' differential imaging (CDI). The idea here is to use the control system or part of the coronagraph itself to provide a known modulation, such as phase probes with the deformable mirror\cite{2006ApJ...638..488B,2007SPIE.6691E...7G}, fringes through controlled leaks at the Lyot stop level (such as with the self-coherent camera\cite{2006dies.conf..553B}), or even modulation of the pupil shape\cite{2008OExpr..1615553R}. Indeed, as rightfully argued in Ref.~\citenum{2006ApJ...638..488B}: focal plane detection coupled with common-path use of the self-coherent nature of each stellar photon through, e.g., successive images, plus a knowledge of the perturbations applied, should allow one to remove most of the speckle field in post-processing.

\subsection{Data reduction}
\label{subsec:fourpillars}  

Data reduction constitutes our fourth ``pillar'', but in reality encompasses the other three. Until the advent of high contrast imaging, astrophysical imaging data reduction long remained basic and mostly cosmetic: dark/background subtraction, flat-fielding, bad pixel and cosmic ray removal. The spreading of AO-corrected imaging triggered the development of more advanced image processing methods (see, e.g.~Ref.~\citenum{2004JOSAA..21.1841M} and references therein), but these did not become a must-have, because astrophysical interpretation from the basically reduced data was still possible.

With the advent of high contrast imaging instrument the perspective is changing quite radically, as the task at hand, namely the detection of a faint signal in noisy background with a dramatic dynamical range from a very large dataset, is simply impossible without suitably designed processing methods. These methods aim at exploiting more and more of the information
available from the instrument and the refined observing strategies devised for this endeavor.

The key before applying any detection criteria is the process of reducing the presence of information not relevant to detection, or in other words calibrating the systematics out. In high contrast imaging, systematics mostly occur in the form of quasi-static speckles generated by slowly varying optical aberrations not seen by the various sensing devices of the instruments (e.g.~the non-common path aberrations). At small angles, as seen in Sect.~\ref{subsec:wfs}, quasi-static speckles morph into more dynamic systematics originating from intricate physical processes, most of the time due to the conjugate action of residual turbulence, temperature variations, and vibrations.

The new generation data reduction algorithms make heavy use of the known modulation of the signal imprinted by the observing strategy employed. It is vital here to distinguish between the strategy and the post-processing. Angular differential imaging is an observing strategy, not a post-processing method. Several post-processing algorithms can be applied to angular differential imaging data. The same goes for spectral differential imaging, or dual-beam imaging. Table~\ref{table:red} summarizes the most used data reduction methods so far.

\begin{table*}[!t]
\begin{center}\caption{Data reduction methods. cADI=classical ADI. rADI=radial ADI. LOCI=locally optimized combination of images. Same nomenclature as in Ref.~\citenum{Lafreniere2007}.\label{table:red}}
\begin{tabular}{|l|c|c|c|c|c|c|}
\hline
Parameter			&Diff. &cADI	&rADI		&LOCI\cite{Lafreniere2007}  	&Spect.~dec.\cite{2007MNRAS.378.1229T} &ANDROMEDA\cite{2008SPIE.7015E..60M,Mugnier-a-09,Cornia-p-10}\\ \hline
\# ref.~frames		&1		     &all avail	&10-100		&10-100	 &$\simeq 10$ &1 (pairs)\\
Excl.~($\lambda/d$)	&N/A		    &N/A		&0.5-1.5		&0.5-1.5	 &1-2 &0.5-1.5\\
Ref.~frame sel.			&$\Delta \lambda$, $\Delta \theta$	   &N/A		&$\Delta t$ 	&Max corr. &$\Delta \lambda$ &N/A\\
Na					&Full FoV &Full FoV	&Full FoV		&10-1000	 &N/A &Pix-res \\
g					&N/A &N/A		&N/A			&0.5-2	 &N/A &N/A\\ 
Coeff.~det. 		&$\chi^2$ 	&median	&median		&SVD(+d)\cite{Pueyo2012} &mean/med./fit &N/A\\ 
Mask zone s			&N/A		&no		&no/yes			&no/yes		  &N/A    &N/A\\
\hline
\end{tabular}
\end{center}
\end{table*}

The Locally Optimized Combination of Images (LOCI\cite{Lafreniere2007}) has proven to be a very powerful method, and is now considered a standard. However, due to obvious geometrical constraints, LOCI is not well adapted to small angles. Indeed, the room available to construct the ``optimization zone'' where the coefficients for the linear optimization are calculated shrinks with the IWA. Two adverse effects are then occurring. Either the size of the zone is chosen accordingly and has to dramatically shrink from a few hundreds resolution elements down to only a dozen or so, provoking inevitable flux losses (which can be calibrated out by injecting fake companions), or the user chooses to keep a constant optimization zone, but is then obliged to reach further out by deforming its nominal squarish geometry to a radially elongated one. In that case, the determination of the coefficient becomes strongly influenced by outer speckles which, as hinted in Fig.~\ref{fig4}, have a completely different behavior and variability, simply because they are generated by totally different physical phenomenon. Both choices are sub-optimal, making LOCI still useful but ill-equipped for small angles.

One promising solution for small-angle data reduction (and corresponding strategy) is the smart use of the telemetric data available from the low-order sensors and/or the science camera itself\cite{2011PASP..123.1434V,2012A&A...539A.126M}. Indeed, using this information, and a comprehensive calibration of the instrument through measurements and/or propagation model, one can reconstruct the shape of the coronagraphic point spread function at any time. This information can be used to discriminate between instrumental effects and real off-axis objects, as demonstrated in Ref.~\citenum{2010sf2a.conf...97Y,2011PASP..123.1434V,2012A&A...539A.126M}.

At least two families of approaches can be considered for detecting exoplanets in speckle noise: 
\begin{enumerate}
\item One is to disentangle the speckle field and the planet in the data by considering both of them as quantities to be estimated jointly using their different behavior, with respect to wavelength in particular (see, e.g., Refs.~\citenum{2007MNRAS.378.1229T,2010sf2a.conf...97Y,Ygouf-p-11b}).
\item Another one to consider the speckles as correlated noise, to pre-process the data so as to remove the correlations in the noise (this is called \emph{pre-whitening}), and then to perform detection.
\end{enumerate}

A unifying framework for the latter approach is detection theory, which shows that the function that optimally discriminates between the planet present and planet absent hypotheses, called the optimal discriminant function, is the ratio of the \emph{likelihood} of the data under the two assumptions. A computationally tractable variant of this function is the Hotelling observer,
which is the optimal \emph{linear} discriminant\cite{Barrett-a-98,2006JOSAA..23.3080B}. For more details about this subject, the reader is invited to read Ref.~Lawson et al.~2012, these proceedings.

LOCI and the likelihood-based ANDROMEDA method can be seen as the first two steps in that direction. LOCI consists in an efficient pre-whitening method of the speckle noise, as the autocorrelation matrix inverted in this algorithm is a simplified analog to the covariance matrix used in the Hotelling observer formalism. ANDROMEDA pre-whitens the noise in a somewhat simplified way by pair-wise image differences and then introduces the rigorous statistical tools of
estimation and detection theory to the subsequent detection task. 

{\bf Last minute update}: right before submitting this review, a new potentially game-changing method was proposed by Soummer, Pueyo, and Larkin\cite{2012arXiv1207.4197S}. The new method is based on the well-known Principal Component Analysis (PCA): assuming a library of reference PSFs, a Karhunen-Lo\`eve transform of these references is used to create an orthogonal basis of eigenimages, on which the science target is projected to create the reference PSF. A PSF constructed in this fashion minimizes the expected value of the least-squares distance between the ensemble of reference images and the random realization of the telescope response contained in the science image. Using projections provides a linear framework that enables forward modeling of astrophysical sources by fitting directly an astrophysical model to the PSF-subtracted data, without introducing degeneracies. PCA, contrary to LOCI seems particularly well adapted to small angles because it does not require the introduction of ``optimization and subtraction zones\cite{Lafreniere2007}'' (see above) and can be modified to accommodate for external telemetry. This new technique has the potential of being another major step towards the unified framework mentioned above.

\section{Lessons learned from the first generation instruments after 10+ years of operations}
\label{sec:firstgen}  

\begin{table}[h]
\caption{Table listing (non exhaustive) the zeroth and first generation instruments that pioneered high contrast imaging. For ground-based near-infrared instruments, the first acronym is usually for the adaptive optics system, while the second one is for the camera. The type of coronagraph is given in the last column, when available. Note that the contrast performance of these instruments varies a lot, depending on the instrument design itself and on the observing strategy. Instruments marked with a \dag\ are no longer available. SH stands for Shack-Hartmann WFS. C for curvature WFS.}
\centering \label{tab4}
\begin{tabular}{|l|l|l|l|l|l|l|}
\hline
Instrument & Telescope &AO & Wavelength   & Ang.\ res. & Coronagraph \\
              &      &      & ($\mu$m)  &   (mas)    &    \\
\hline
WFPC2\dag     & HST  &NA      & 0.12--1.1 & 10--100  & ... \\
NICMOS\dag    & HST &NA       & 0.8--2.4  & 60--200  & Lyot\\
ACS           & HST  &NA      & 0.2--1.1  & 20--100  & Lyot\\
STIS          & HST   &NA     & 0.2--0.8  & 20--60   & Lyot\\
NAOS-CONICA   & VLT &16-SH       & 1.1--3.5  & 30--90   & Lyot/FQPM/APP(/OVC)\\
VISIR         & VLT    &no    & 8.5--20   & 200--500 &FQPM/OVC \\
COME-ON+-ADONIS &3.6-m ESO &8-SH &1--5 &60--280 &Lyot \\ 
PUEO-TRIDENT &CFHT &8-SH  &0.7--2.5 &4--140 &Lyot(/CIA) \\
HICIAO          & Subaru &14-C    & 1.1--2.5  & 30--70   &Lyot/PIAA/8OPM\\
AO-NIRC2      & Keck   &16-SH    & 0.9--5.0  & 20--100  & Lyot\\
LWS           & Keck  &no     & 3.5--25   & 70--500  & ...\\
MIRLIN\dag    & Keck   &no    & 8.0--20   & 160--400 & ...\\
ALTAIR-NIRI   & Gemini N. &no  & 1.1--2.5  & 30--70   & Lyot\\ 
NICI          & Gemini S. &9-C & 1.1--2.5  & 30--70   & Lyot\\
T-ReCS        & Gemini S. &no & 1.1--2.5  & 30--70   & ...\\
Lyot project\dag  & AEOS &30-SH  & 0.8--2.5  & 60--140  & Lyot/FQPM\\
PALAO(WCS)-PHARO\dag   & Hale 200'' &16-SH & 1.1--2.5  & 60--140  & Lyot/FQPM/OVC\\
P3K-P1640/PHARO     & Hale 200'' &64-SH & 1.1--2.5  & 300  & APLC/OVC\\
AO-IRCAL      & Shane 120'' &8-SH & 1.1--2.5  & 100--150 & ... \\
PISCES      & LBT' &30-Pyramid & 1.0--2.5  & 30--60 & ... \\
LMIRCAM  & LBT' &30-Pyramid & 2--5  & 60--120 & ...\\
\hline
\end{tabular}
\end{table}

The first (and zeroth) generation instruments capable of providing good quality wavefronts (hence image) were truly venturing into a whole new domain. They were pioneers, advancing instrumentation, and introducing adaptive optics to mainstream astronomy.  Table~\ref{tab4} presents a census of past and current facilities, outlining their wavelength range, angular resolution and coronagraph capabilities. Note that most of these instruments were multi-purpose instruments and were not optimized for high contrast imaging. One obvious observation is that adaptive optics has undergone a quite recent outbreak, and that Lyot coronagraphs are widespread. We note that new generation \emph{small-angle} coronagraphs (FQPM, OVC, PIAA, etc.) are making their way into current generation facilities, most of the time as limited but successful experiments.

The lessons learned from the first generation of high-contrast imagers (adaptive optically assisted or not) can be summarized in three main points, which were raised and discussed during the Exoplanet Imaging Workshop held 25-30 March 2012 at the Resort at Squaw Creek, California, and whose participants are co-authoring this review:
\begin{enumerate}
\item Instrument knowledge and operations: it is critical to transfer the knowledge from the instrument building teams to the operation teams and users. Given the complexity of high contrast imagers (especially when adaptively corrected), it is essential to allow users and maintenance teams to keep the instrument in the best shape possible over its lifetime, in order to ensure continuous optimal performance delivery. Keeping a healthy level of activities and upgrades is also vital for the motivation, and most of the time allows unintended technical and scientific breakthroughs.   
\item Stability: calibration non-common path errors are key to the final performance levels in high contrast imaging. Non-common path errors can themselves be classified into three categories, purely static (permanent defects in optics), quasi-static (temperature and gravity-vector induced slowly varying aberrations), and dynamic (vibrations). The importance of the latter was probably heavily underestimated by both instrument building teams and observatories. These need to be better characterized, control, and observatory-wide policies need to be put in place so that the nominal performance level of the instrument environment (platform, telescope, dome, etc.) can be guaranteed.
\item Detection of faint signal in high contrast imaging must rely on diversity and modulation: it indeed appeared early on that calibration and correction of spurious errors is not sufficient for detecting faint signals against the background of noise sources. Modulation tremendously helped with detection and now constitutes the baseline for any strategy (see Section~\ref{subsec:strategy}).
\end{enumerate}

Other important considerations include: flexibility vs science, technical development, funding, chromatism effects, coronagraph design, ADC, propagation effects, importance of predictive control, image quality metrics, and error budgets, actuator count vs outer working angle (OWA), DM stability, etc.

\section{Choices made for second generation instruments}
\label{sec:secondgen}  

It is striking how diverse the second generation of high contrast imagers are despite the breakthrough fact that for the first time, these instruments are fully optimized to perform high contrast imaging. Indeed, while all focused on the exact same science goals (and targets!), the technical choices made by each project are covering the range of technologies currently available. This accidental cross-breeding is an excellent thing. It is expected that ``technical and scienfitic selection'' will then decide which technology is the most adapted to a particular range of scientific goals (e.g.~detection vs characterization). For this reason, the experience of the second generation instruments is vital for the future of high contrast imaging and the long term goal of Earth-like/unlike planet detection in space and/or from the ground with giant apertures.

\subsection{Different philosophies}
\label{sec:secondgen} 
GPI\cite{2008SPIE.7015E..31M} and SPHERE\cite{2006Msngr.125...29B} have chosen two very different technological solutions, mostly dictated by the fact that SPHERE will be sitting on the big Nasmyth platform of the VLT UT3, while GPI will be attached to the Cassegrain focus of Gemini South. In many respects, the facility and telescope had a big influence on the design. GPI is more compact and simplified than SPHERE. Optics are much smaller on GPI, including the MEMS DM, which could not be made with standard piezo-stack technology like SPHERE. Also because the instrument is moving with the telescope constantly, a dedicated calibration system is built in GPI. This CAL system integrates the measurement of low-order and higher-order aberrations in quasi real time (on timescales of minutes). On the other hand, SPHERE relies on its stability through temperature and vibration control, while counting of classical static calibration through well known phase diversity techniques, currently successfully implemented and routinely used at major observatories on weekly-daily or even hourly basis: Keck-NIRC2's image sharpening routine, NACO's phase diversity \cite{2003A&A...399..385H}, Palomar-PHARO's modified Gerchberg-Saxton \cite{2010SPIE.7736E.197B}, etc.

Note  that P3K-P1640 at Palomar relies on a CAL system similar to that of GPI\cite{2011PASP..123...74H}. ScEXAO at Subaru\cite{2011SPIE.8149E...7G} uses a very original solution, based on the dedicated CLOWFS, paired with its PIAA coronagraph. 

\subsection{Lessons already learned from systems in development and about to be commissioned}
Even though most of second generation instruments have not been fully commissioned yet\footnote{Some of them, like Palomar's P3K (Roberts et al.~2012, these proceedings) and Subaru's ScEXAO (Guyon et al.~2012, these proceedings) have already been tested on sky, to different levels.}, we can already draw a few conclusions of the 10 years or so of theoretical design, subsystems manufacturing and testing, and integration, which is taking place right now. Again, let us put forward three main points, which were raised and discussed during the Exoplanet Imaging Workshop held 25-30 March 2012 at the Resort at Squaw Creek, California:
\begin{enumerate}
\item Importance of identifying critical hardware items, e.g.~detectors (especially the visible sensor detector, which must be fast and present low noise properties), deformable mirrors (absolutely critical for second generation instrument), etc. Given the complexity and the uneven technology readiness level of the various subsystems of these new complex instruments, the planning needs to allow for reserves in schedule and cost. It is also important to accommodate for the flexibility of including new and better technologies as they come along during the development phase of the instrument (and eventually during its integration, commissioning and lifetime!). One excellent example of such a flexibility is the consensus now in the community of the inadequacy of the Shack-Hartmann wavefront sensor for small-angle high contrast imaging. Recent outstanding results and routine performance levels of the Pyramid WFS of the LBT is the real-world demonstration of the superiority of diffraction-limited WFS\cite{2005ApJ...629..592G}.
\item Importance of optical quality and propagation effects. When reaching to $10^{-6}$ contrast levels and below, frequency folding\cite{2004SPIE.5490.1438G} and Talbot effects start to kick in very significantly, affecting the chromatic behavior of speckles in such an important way that it invalidates the basic principles of spectroscopic observing strategies and associated data reductions algorithms (e.g.~spectral deconvolution\cite{2007MNRAS.378.1229T}). This diffraction effect was probably underestimated by most instrument building teams. Research in high contrast imaging for space-based projects, because it reaches to $10^{-9}-10^{-10}$ contrast levels, has learned this lesson a long time ago and now fully integrates Fresnel propagation effects into wavefront control methodologies and algorithms\cite{2007SPIE.6691E...7G}. Second generation projects might not have taken the low hanging fruits from this research soon enough. Moreover, propagation effects emphasize or even sometimes amplify the adverse effects of optical defects (wavefront quality, coating uniformity, cleanliness), imperfection of ADC designs and ill-advised choice of their location. 
\item Somewhat related to the previous point, new approaches to wavefront control (again inherited from space-based coronagraph research\cite{2007SPIE.6691E...7G}) have been integrated to the new projects at their very end. But it is now a consensus that focal plane calibration and image wavefront control plays an essential part in the crucial calibration of static errors. There is room for additional improvements, such that two-DMs designs for $360^\circ$ control of phase and amplitude\cite{2011arXiv1111.5111P} (as opposed to one-sided dark holes). In atmospheric turbulence correction, predictive control \cite{2008JOSAA..25.1486P} is key to improving servo-lag limitations on guide star magnitude and subaperture size.
\end{enumerate}

Other lessons currently being learned include: image quality metrics (obsoleteness of the Strehl ratio), and error budgets (naive view of the usual ``spreadsheet quadratic sum''), actuator count vs. OWA, DM stability, huge importance of detectors (RON, speed, calibration accuracy), spectral bandwidth i.e.~what can be simultaneously supported, coronagraph choice and designs for small-IWA systems, investigate optimal spectral resolution: trade spectral resolution against FOV, etc.

\section{State-of-the-art and recent progress}
\label{sec:stateoftheart}  

\begin{figure}[h]
\centerline{\includegraphics[width=17cm]{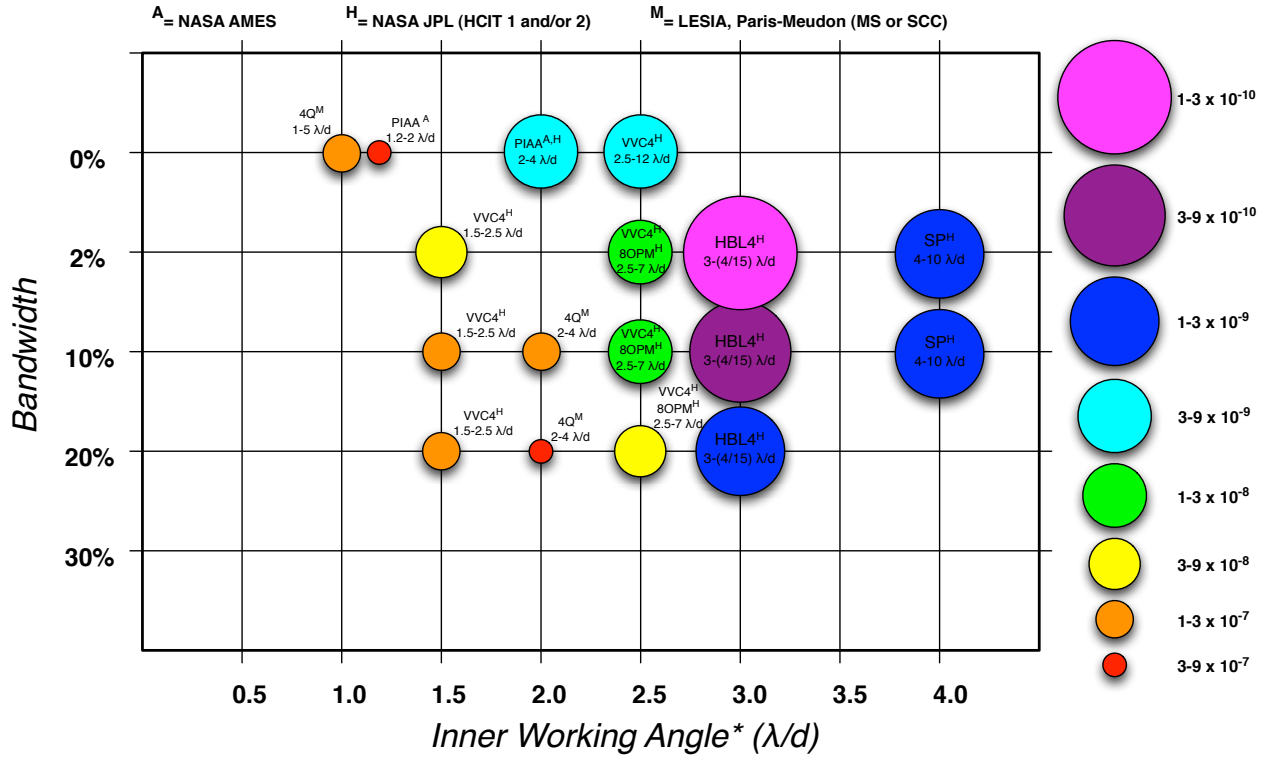}}
\caption{Raw contrast world records for small-angle coronagraphs as a function of spectral bandwidth and inner working angles. The IWA used in this graph differs from the universal definition presented above, and restricted to the pure coronagraph capabilities. Here, the IWA is defined as the one set arbitrarily by the combination of the coronagraph and the wavefront control system. We purposefully limited ourselves to the coronagraph allowing IWA smaller than $4\lambda/d$.
\label{fig7}}
\end{figure}

The state of the art in small-angle coronagraphy and associated wavefront control techniques can undoubtedly be found in lab experiments dedicated to the demonstration of the highest contrast coronagraphic technologies for space-based mission concepts. Several facilities are dedicated to that purpose in Hawaii, NASA JPL, NASA Ames and at Paris-Meudon observatory. The highest raw contrast at the smallest angles and larger bandwidths have however been obtained at JPL, where the so-called High Contrast Imaging Testbed (HCIT) provides the ideal vacuum environment for such technological demonstrations (see Ref.~\citenum{2007SPIE.6693E..27T} for details). 

Using a reflective scheme, this facility highlights a vacuum chamber mounted on vibration isolation blocks, and high-order  deformable mirrors (DM) to perform accurate wavefront control. Two types of DM are currently available: a new superpolished 64x64 Xinetics mirror, and a series of 32x32 Xinetics mirrors. The HCIT available light sources (lasers, supercontinuum white light sources) can simulate a point source across a variety of wavelengths and bandpasses in the visible range.

It is interesting to note that, using a modified Gerchberg-Saxton estimation procedure very similar to what is currently done on ground-based telescopes and foreseen for second-generation instruments, the technique typically leaves a 5 nm rms residual Òfront-endÓ wavefront error, mostly outside of the control spatial bandwidth (extending to $N/2\times \lambda/d$, where N is the linear number of available actuators across the pupil). The initial contrast with this corrected phase is typically limited to a few $10^{-6}$ within the control region, because of frequency folding and chromatic phase to amplitude errors conversion due to propagation effects\cite{2004SPIE.5490..545P,2004SPIE.5490.1438G,2006AAS...20916410P,2006ApOpt..45.5143S}.

To enhance contrast further and create a so-called dark hole, i.e.~a region of the image where the wavefront correction provided by the DM is computed, the electric field conjugation algorithm, or EFC \cite{2007SPIE.6691E...7G} is now routinely used. The name of the algorithm comes from the fact that the DMÕs actuators are set so as to superpose the negative of the electric field onto the image plane, conceptually making the image intensity zero (more rigorously, minimizing energy). The algorithm is divided into two parts, estimation and correction. The complex electric field amplitude in the image plane is estimated, then the estimate is processed to determine a correction to be applied to the DM actuators. These two parts are functionally independent, and each could be used separately with other approaches. Each EFC iteration begins with a DM setting which is the result of the prior iterationÕs correction result (or is the flat setting, to begin the first iteration). 

While showcasing mostly HCIT results, Fig.~\ref{fig7} also presents results obtained on the NASA AMES high contrast testbed, which features Boston micromachine MEMS DMs and is dedicated to the PIAA coronagraph\cite{2011SPIE.8151E...1B}. The AMES testbed is not operated under a vacuum environment but benefits from sophisticated temperature, air flux control and acoustic isolation. Fig.~\ref{fig7} also shows results obtained at LESIA (Paris-Meudon observatory), where several testbeds dedicated to the demonstration of high contrast coronagraphy have been built over the years. These testbeds are also in air, and benefit from various degrees of temperature and turbulence control\cite{2011A&A...530A..43G,2012A&A...539A.126M}.

\section{Conclusions}
\label{sec:conclusions}  

This paper has reviewed the current state-of-the-art technologies and techniques used for high contrast imaging at very small angles. The review is timely since ground-based second-generation adaptive optics systems dedicated to the detection and characterization of extra-solar planets are currently coming online. One of the important conclusions of this review is the under-estimated potential of the scientific and technical cross-fertilization between space-based and ground-based projects. Indeed, even if science cases and technologies must stay optimized and therefore specific, there are many lessons that both endeavors can learn from each other. Because of the much tighter requirements set by the very challenging goal of Earth-like planet imaging, space-based concepts are pushing technologies forward, and are routinely confronted to physical limits that the ground-based projects are only starting to learn while catching up in terms of contrast levels. On the other hand, space-based mission concepts can benefit from the extensive operational experience of the 10+ years of adaptive optics operations on ground-based telescopes. Indeed, while having to cope with stringent limitations from the atmosphere and usually bad optics, the ground-based community has developed unique strategies (observing and post-processing) that are well proven, and could be studied for space-based coronagraphs. In all aspects reviewed in this paper, the lessons from the commissioning and operation of the second generation instruments will be crucial for the future of high contrast imaging in space and on the ground.

\acknowledgments     
This work was carried out at the European Southern Observatory (ESO) site of Vitacura (Santiago, Chile), and the Jet Propulsion Laboratory (JPL), California Institute of Technology (Caltech), under contract with the National Aeronautics and Space Administration (NASA). The first author is grateful to the organizers and participants of the Exoplanet Imaging Workshop held 25-30 March 2012 at the Resort at Squaw Creek, California.


\bibliography{mawet_spie2012a_v2}   
\bibliographystyle{spiebib}   

\end{document}